\documentclass[amsmath,amssymb,aps,twocolumn,prl,floatfix]{revtex4}

\usepackage{mathrsfs}
\usepackage{amsmath}
\usepackage{amssymb}
\usepackage{color}
\usepackage{graphicx}	
\usepackage{bm}
\usepackage{ulem} 
\usepackage{titlesec}
\titleformat{\paragraph}[runin]
{\bfseries\scshape}{\theparagraph}{1em}{}
\usepackage{braket}
\usepackage{multirow}

\usepackage[printwatermark]{xwatermark}
\usepackage{xcolor}

\newcommand{\be}{\begin{equation}}
\newcommand{\ee}{\end{equation}}
\newcommand{\bef}{\begin{figure}}
\newcommand{\eef}{\end{figure}}
\newcommand{\bea}{\begin{eqnarray}}
\newcommand{\eea}{\end{eqnarray}}

\newcommand{\brho}{\mbox{\boldmath${\rho}$}}  
\DeclareMathOperator\erf{erf}
	     
\begin{document}
\title{Computing Transition Rates for Rare Event: When Kramers Theory meets Free Energy Landscape}
\author{Fran\c cois Sicard}
\thanks{Corresponding author: \texttt{francois.sicard@free.fr}.}
\affiliation{Department of Chemical Engineering, University College London, Torrington Place, London WC1E 7JE, United Kingdom, EU}

%
\begin{abstract}
Computing reactive trajectories and free energy (FE) landscapes associated to rare event kinetics is key 
to understanding the dynamics of complex systems. The analysis of the FE surface on which the underlying dynamics takes place 
has become central to compute transition rates.
In the overdamped limit, most often encountered in biophysics and soft condensed matter, the 
Kramers' Theory (KT) has proved to be quite successful in recovering correct kinetics. 
However, the additional calculation to obtain rate constants in complex systems where configurational entropy 
is competing with energy is still challenging conceptually and computationally.
Building on KT and the metadynamics framework, the rate is expressed in terms of the height of the FE barrier 
measured along the minimum FE path and an auxiliary measure of the configurational entropy.
We apply the formalism to two different problems where our approach shows good agreement with simulations 
and experiments and can present significant improvement over the \textit{standard} KT.
\end{abstract}

\maketitle

Since the seminal work of Hendrik A. Kramers in 1940~\cite{1940-Physica-Kramers}, the study of rare events 
has been a subject of considerable interest to several scientific communities~\cite{1990-RMP-Hanggi-Borkovec,
2010-ARPC-Weinan-Vanden,2010-JCP-Xin-Hamelberg,2012-PRL-Gobbo-Baroni,2013-PRL-Tiwary-Parrinello,
2014-JCTC-Salvalaglio-Parrinello,2015-JCP-Sicard-Manghi,2016-JCP-Mokkonen-Jonsson,2017-RP-Pietrucci}. These events are rare because the systems of interest have to overcome some barriers, 
which can either be of an energetic or an entropic nature. 
From a theoretical viewpoint, rate theories, such as transition-state theory~\cite{1996-JPC-Truhlar-Klippenstein} (TST) 
and Kramers' theory~\cite{1940-Physica-Kramers,1990-RMP-Hanggi-Borkovec} (KT), 
have been successful in providing the language, the intuition, and the foundation for the development of 
computational tools for studying barrier-crossing events. What is most attractive  about rate theory is its simplicity.
It states basically that to move from the \textit{reactant} state to the \textit{product} state, the system has to navigate 
itself to the transition state, which is a saddle point on the potential, or free energy (FE) surface. 
In many cases, one can also define the most probable transition path for the reaction, which for overdamped 
systems of interest here is simply the minimum FE path (MFEP).

Molecular dynamics (MD) simulations are now used on a regular basis to study the statistical properties 
of barrier-crossing events in the long-time limit~\cite{2010-JCP-Xin-Hamelberg,2012-PRL-Gobbo-Baroni,
2013-PRL-Tiwary-Parrinello,2014-JCTC-Salvalaglio-Parrinello,2015-JCP-Sicard-Manghi}. 
In the context of rare events, the systems can present different FE minima, each one trapping the dynamics 
for a time that can be long compared to fast bond vibrations, until a thermally activated jump is eventually performed toward 
another metastable or global minima. 
Ideally, a complete understanding of an activated process would encompass all of its kinetic 
aspects. However, there is often a wide gap between the time scale of the transition of interest 
and the time scale accessible with simulations, 
and one is content with reconstructing the geometric pathways and their FE profiles. 
To do so, a number of different computational approaches were introduced in the last few decades, 
sometimes designed on purpose and sometimes borrowed from different disciplines~\cite{2017-RP-Pietrucci}. 
Nevertheless, it remains necessary to asses the reliability of these methods with comparison with appropriate 
rate theory~\cite{2017-Elsevier-Peters}.
%

In the present work, we consider the overdamped limit most often encountered in biophysics and soft condensed 
matter~\cite{2015-JCP-Sicard-Manghi,2017-arXiv-De-Gupta}, for which the KT has proved to be 
quite successful in recovering correct kinetics.
Focusing on complex systems characterized with metastable states where entropy is competing with energy, 
we introduce a new approach to evaluate transition rates when configurational entropy~\cite{1981-Macromolecules-Karplus-Kushick,
2012-PCCP-Nguyen-Derreumaux,2017-arXiv-De-Gupta} associated to anharmonic motions in the metastable basin 
and not captured by the MFEP comes into play. Building on the \textit{standard} KT and 
the metadynamics~\cite{2002-PNAS-Laio-Parrinello,2008-RPP-Laio-Gervasio} (metaD) framework, the rate is first expressed 
in terms of the height of the FE barrier measured along the MFEP. We then define an auxiliary measure of 
the configurational entropy in the metastable basin based on the reconstruction of the FE 
landscape obtained from metaD simulations~\cite{2017-JCP-Gimondi-Salvalaglio}.\\

The starting point in the theory of barrier crossing under the influence of friction initiated by Kramers 
is the inertial Langevin equation with Markovian friction and random forces coupled to 
reaction coordinate motion~\cite{2012-WS-Coffey-Kalmykov}:
\begin{equation}
m\ddot{q} = -\frac{\partial V}{\partial q} - \gamma \dot{q} + R(t) \,.
\label{Langevin}
\end{equation}
In Eq.~\ref{Langevin}, $q$ represents the reaction coordinate, $m$ is the reduced mass for the reaction coordinate, 
$\gamma$ is the friction coefficient, and $V(q)$ is a potential of mean force (PMF). $R(t)$ is a random force with 
zero mean that satisfies the fluctuation-dissipation theorem~\cite{2008-PR-Marini-Vulpiani}. 
Without loss of generality, we set $m = 1$ in the following.
In principle, Langevin equation can be constructed from MD simulations. For instance, 
the PMF can be computed using metaD or umbrella sampling simulations.
KT is a valid approximation for real solvent as encountered in polymer physics and classic theories of nucleation 
and provides a unified framework for understanding how dynamics influence reaction rates~\cite{2017-Elsevier-Peters}. 
In particular, the strong friction limit of interest here is where quantitative results from KT are most reliable. 
In this limit, the time evolution of the probability density $P(x,t)$ is governed by the Smoluchowski equation~\cite{1940-Physica-Kramers}
\begin{equation}
\frac{\partial P(q,t)}{\partial t} = -\frac{1}{\gamma}\frac{\partial}{\partial q} \Big(\frac{\partial V}{\partial q}P(q,t)+\frac{1}{\beta}\frac{\partial P(q,t)}{\partial q}\Big)\,,
\label{Smoluchowski}
\end{equation}
where the right-end term in Eq.~\ref{Smoluchowski} corresponds to the gradient of the probability flux $J$ 
over the barrier 
\begin{equation}
J = -\frac{1}{\gamma} e^{-\frac{V(q)}{k_B T}} \frac{\partial}{\partial q}\Big( e^{\frac{V(q)}{k_B T}} P(q,t) \Big)\,,
\label{flux}
\end{equation}
considering the system is thermalized near the bottom of the well~\cite{1940-Physica-Kramers}.
Following the original reasoning of Kramers~\cite{1940-Physica-Kramers}, we assume a steady state escape rate, $k_{KT}$, 
by considering a stationary situation for the the probability flux $J$, $\frac{\partial P}{\partial t}=0$. For sufficiently 
high FE barrier the probability density follows the equilibrium relation $P(q) = P(q_0) \exp\big(-(V(q)-V(q_0))/k_B T\big)$.
Integrating Eq.~\ref{flux} along the PMF and expanding about the transition state, $q_{T}$, yields
\begin{equation}
J = P(q_0) \frac{\sqrt{\lvert V''(q_{T})\rvert}}{2\pi\gamma} e^{-\frac{V(q_{T})-V(q_0)}{k_B T}}\,.
\label{final-flux}
\end{equation}
Rewriting $J=p~k_{KT}$, with $p$ the probability of the particle being inside the metastable well
and $k_{KT}$ the Kramers' escape rate, we consider 
that the system is confined to a small neighbourhood $\Omega_{q_0}$ around the minimum $q_0$ of the well. 
Expanding about this point, the probability of finding a particle in the well is
\begin{equation}
p = \int_{\Omega_{q_0}} P(x) dx = P(q_0) \sqrt{\frac{2\pi k_B T}{V''(q_0)}}\,.
\label{}
\end{equation}
This yields the Kramers' escape rate, 
\begin{equation}
k_{KT} =  \frac{\sqrt{V''(q_0) \times \lvert V''(q_T)\rvert}}{2\pi \gamma} e^{-\Delta V/k_B T}\,,
\label{KramersEQ}
\end{equation}
where $\Delta V = V(q_{T})-V(q_0)$. The expression in Eq.~\ref{KramersEQ} must account for the symmetric or asymmetric 
nature of the FE profile 
in the metastable states and at the transition state. To do so, the PMF $V(q)$ in Eq.~\ref{Langevin} can either be fitted 
with Gaussian or skew-Gaussian curve depending on the symmetric or asymmetric nature of 
the FE profile~\cite{2014-BJ-Woodside-Beach,2017-arXiv-De-Gupta}, respectively
\begin{eqnarray}
V_{\textrm{sym}}(q) &\propto & e^{-(q-q_0)^2/2\sigma^2}\,, \label{symPot}\\
V_{\textrm{asym}}(q) &\propto & V_{\textrm{sym}}(q)\Bigg(1+\textrm{erf} \Big(\frac{\alpha (q-q_0)}{\sqrt{2}\sigma}\Big) \Bigg)\,,\label{asymPot}
\end{eqnarray}
with $\sigma$ and $\alpha$ the parameters of the distributions. We can then rewrite Eq.~\ref{KramersEQ} 
in the form of the expression originally derived by Kramers in the overdamped regime~\cite{1940-Physica-Kramers}, 
\begin{equation}
\label{KramersEQ-omega}
k_{KT} = \frac{\omega^{eff}_0 \omega^{eff}_{T}}{2\pi \gamma} e^{-\Delta V/k_B T} \,,
\end{equation}
where $\omega^{eff}_0$ and $\omega^{eff}_{T}$ represent the \textit{effective} stiffness of the well and the barrier, 
respectively, modeled with the Gaussian or skew-Gaussian distributions in Eqs.~\ref{symPot}-\ref{asymPot}\\

The KT discussed above gives a physical derivation of the reaction rate constant, $k_{KT}$, in terms of 
the shape of the FE profile. 
This consideration comes closer to reality for a reaction with a FE landscape containing a large energy barrier 
and narrow valley between \textit{reactants} and \textit{products}, but it will be a poor approximation in the presence of 
\textit{large} entropic FE basins~\cite{2017-Elsevier-Peters}. In such case, the convergence of the FE profile 
could even not be achieved due to \textit{large} entropic fluctuations.
To overcome this limitation, we consider the shape of the MFEP in Eq.~\ref{KramersEQ-omega} instead. 
As the transition rate defined in Eq.~\ref{KramersEQ-omega} may account for the activation entropy 
captured by the MFEP, such as rotational and vibrational entropy~\cite{2013-JACS-Flaherty-Iglesia,2017-ACR-Aqvist-Brandsdal,
2017-JCP-Sensale-Chang}, it does not account for the configurational entropy, $S^{\textrm{conf}}$, 
in the metastable basin~\cite{2007-CRC-Leuzzi-Nieuwenhuizen,2012-PCCP-Nguyen-Derreumaux}:
\begin{equation}
\label{Sconf}
 S^{\textrm{conf}} = -k_B \int dx \rho(x) \ln \rho(x) \,,
\end{equation}
where $\rho(x)$ represents the canonical probability density distribution function associated with 
the system potential energy $U(x)$ of the form
\begin{equation}
\label{PDFconf}
 \rho(x) = \frac{\exp(-U(x)/k_BT)}{\int dx \exp(-U(x)/k_BT)}   \,.
\end{equation}
Since its introduction in 1981 by Kushick and Karplus in the context of macromolecules~\cite{1981-Macromolecules-Karplus-Kushick}, 
a number of methods have been proposed in the literature to estimate the configurational entropy 
of complex systems~\cite{2004-JCP-Peter-vanGunsteren,2006-JCP-Cheluvaraja-Meirovitch,
1981-Macromolecules-Karplus-Kushick,2012-PCCP-Nguyen-Derreumaux}.
%
\begin{figure*}[t]
\includegraphics[width=0.95 \textwidth, angle=-0]{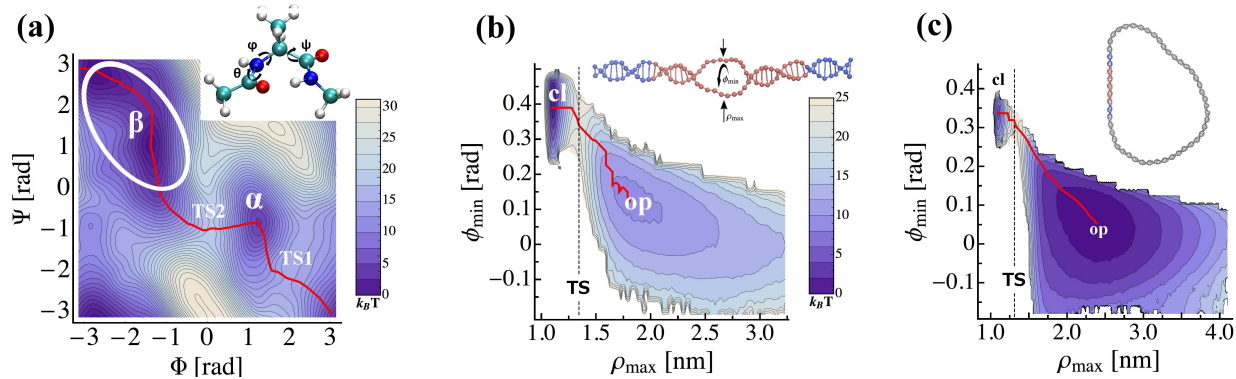}
 \caption{\textbf{(a)} FE surface associated with the conformational transition between conformers 
 $\alpha$ and $\beta$ of alanine dipeptide in vacuum as a function of the two dihedral angles $\Phi$ and $\Psi$ (see inset). 
 The two minima $C_{7eq}$ and $C'_{7eq}$ are combined in the $\beta$ basin as in Refs.~\cite{2000-PNAS-Bolhuis-Chandler,2013-PRL-Tiwary-Parrinello}. The contour lines are every half $k_B T$.
 \textbf{(b)} FE surface associated with the linear DNA bubble closure/nucleation mechanism projected along the maximal 
 distance between paired bases $\rho_{\textrm{max}}$ and the minimal twist angle between successive bps, $\phi_{\textrm{min}}$ 
 (see inset).  The two stables basins associated with the opened (op) and closed (cl) states of the DNA bubble are shown. 
 The contour lines are every two $k_B T$.
 \textbf{(c)} FE surface associated with the circular DNA bubble closure/nucleation mechanism projected along $\rho_{\textrm{max}}$ and 
 $\phi_{\textrm{min}}$. The two stables basins associated with the opened (op) and closed (cl) states of the DNA bubble are shown. 
 The contour lines are every two $k_B T$.
 In the three examples, the typical MFEP obtained within the steepest descent framework~\cite{2013-JCP-Chen-Xiao} are shown in red color. }
\label{fig1}
\end{figure*}
We consider here the definition of the FE difference between two metastable basins $\mathcal{B}_i$ and $\mathcal{B}_j$, 
$\Delta F^{*}_{ij}$, in terms of the probability distribution of the collective variables (CVs) 
along which the FE landscape is projected~\cite{2017-JCP-Gimondi-Salvalaglio} to assess 
quantitatively the entropic contribution of the FE surface:
\begin{equation}
\label{FE-entropy}
\Delta F^{*}_{ij} = -k_B T ~\log \Big( \frac{P_i}{P_j}\Big) \,.
\end{equation}
In Eq.~\ref{FE-entropy}, $P_i$ and $P_j$ are the probabilities of states $i$ and $j$, respectively. 
The probability of each state is computed as the integral of the distribution within the FE basin, 
$\mathcal{B}$, it occupies on the CV-space reconstructed within the metaD framework,
\begin{equation}
\label{Proba-definition}
 P_i = \iint_{\mathcal{B}_\textrm{i}} f({CV_1, CV_2,\dots}) ~dCV_1~dCV_2 \dots  \,,
\end{equation}
where $f$ is the joint probability density distribution function associated with the system FE,  
and $\{CV_1$, $CV_2$, $\dots$\} represents the subset of CVs needed to reach the convergence of the metaD simulation and 
to decipher the configurational entropic contribution to the system. This means that the number of CVs to be considered 
in Eq.~\ref{Proba-definition} can be higher than the one needed for the convergence of the FE landscape.
Denoting $\Delta F_{ij}$ the FE of formation between the two metastable basins $\mathcal{B}_i$ and $\mathcal{B}_j$ 
measured along the MFEP, the difference in configurational entropy, 
$\Delta S^{\textrm{conf}}_{ij}$, would be assessed as~\cite{2007-CRC-Leuzzi-Nieuwenhuizen} 
\begin{equation}
\label{Sconf-metaD}
 -T\Delta S^{\textrm{conf}}_{ij} = \Delta F_{ij} - \Delta F^*_{ij} \,.
\end{equation}
Substituting the FE of formation with the FE of activation between the equilibrium basin $\mathcal{B}_0$ 
and the transition state in Eqs.~\ref{Sconf-metaD} and \ref{FE-entropy}, one can rewrite the Kramers' equation as:
\begin{equation}
\label{KTconf}
 k_0 =  k_{\textrm{conf}} \times k_{KT} = e^{\Delta S^{\textrm{conf}}/k_B} ~ \Big(\frac{\omega^{eff}_0 \omega^{eff}_{T}}{2\pi \gamma} 
 e^{-\Delta V/k_B T}\Big) \,,
\end{equation}
with $k_{\textrm{conf}}$ a correction factor accounting for the difference in configurational entropy between 
the equilibrium basin $\mathcal{B}_0$ and the TS.
In Eq.~\ref{KTconf}, $\Delta V = V(q_T)-V(q_0)$ can either represent a potential energy difference, as originally 
considered by Kramers~\citep{1940-Physica-Kramers}, or a FE difference, as considered thereafter.
Eventually, the direct estimation of the transition rate, $k_0$, can be determined if the reduced mass, $m$, 
and the effective friction coefficient, $\gamma$, defined in Eq.~\ref{Langevin} are known~\cite{2017-Elsevier-Peters}.
However, these parameters might not be easy to determine when the complexity of the CVs increases. 
In addition, care must be taken with the direct estimation of the transition rates derived in Eq.~\ref{KTconf}, as 
it is not true to say that there is a fully established Boltzmann-Gibbs distribution in the neighbourhood of the 
transition state~\cite{1980-Landau-Lifshitz,1983-JPC-Laldler-King}.
Nevertheless, it is possible to compute the ratio between the rates associated to the transition between two metastable 
basins, $\mathcal{B}_i$ and $\mathcal{B}_j$: 
\begin{equation}
\label{KTfinal}
\frac{k_{i}}{k_{j}} = e^{\Delta S^{\textrm{conf}}_{ij}/k_B} ~ \frac{\omega_i}{\omega_j} ~ \frac{\gamma_j}{\gamma_i} 
 ~ e^{-(V_i - V_j)/k_B T} \,.
\end{equation}
In the following, we proceed with three illustrative applications of our approach, 
 each with different level of coarse-graining and entropic contribution. 
 The details of the numerical simulations are given in the Supplemental Material (SM).\\

\textbf{Alanine dipeptide in vacuum.} The conformational transition between conformers $\alpha$ and $\beta$ 
of this peptide has been extensively studied as an example of rare event~\cite{2005-JCP-Ren-Weinan, 2013-PRL-Tiwary-Parrinello,
2014-JCTC-Salvalaglio-Parrinello,2017-JMM-Cuny-Mineva,2018-arXiv-Gimondi-Salvalaglio}. 
We performed well-tempered metaD (WT-metaD) atomistic simulations~\cite{2008-PRL-Barducci-Parrinello,2014-PRL-Dama-Voth} using 
both torsional angle $\Phi$ and $\Psi$ as CVs. 
%
In Fig.~\ref{fig1}a is shown the FE surface for this molecule, along with the rough locations of the stable states.
The location of the metastable basins and the heigh of the FE barriers are in agreement with the ones found 
in the literature~\cite{2005-JCP-Ren-Weinan, 2013-PRL-Tiwary-Parrinello,
2014-JCTC-Salvalaglio-Parrinello}. We determined the value of the FE of formation, 
$\Delta F_{\alpha \beta} = F(\alpha) - F(\beta)\approx 4~k_B T$ 
along the MFEP  depicted in Fig.~\ref{fig1}a 
and the FE of formation 
$\Delta F^*_{\alpha \beta}\approx 5~k_B T$, defined in term of the probability distribution of $\Phi$ and $\Psi$, 
considering the successive isosurfaces depicted in Fig.~\ref{fig1}a as integration domain. 
The exact values are given in the SM along with the values of the parameters 
$\omega_{\alpha}$ and $\omega_{\beta}$ defined in Eq.~\ref{KTfinal}.
Assuming that the effective friction coefficient, $\gamma$, in Eq.~\ref{KTconf} remains unchanged in the transitions 
$\alpha \leftrightarrow \beta$, one obtains the transition rate ratio, 
$k_{\beta \to \alpha}/k_{\alpha \to \beta} = (5.6 \pm 2.0) \times 10^{-2}$. This result is in good agreement with the numerical ratio 
obtained within the accelerated MD framework~\cite{2010-JCP-Xin-Hamelberg,2013-PRL-Tiwary-Parrinello,2014-JCTC-Salvalaglio-Parrinello} $k^{(num)}_{\beta \to \alpha}/k^{(num)}_{\alpha \to \beta} = (4.0 \pm 1.5) \times 10^{-2}$.
For instance, the \textit{standard} KT yields $k^{(st)}_{\beta \to \alpha}/k^{(st)}_{\alpha \to \beta} = (1.4 \pm 0.2) \times 10^{-2}$, 
which does not differ significantly from our approach. 

\textbf{Linear DNA denaturation bubble.} The cooperative opening and closure of a sequence of DNA consecutive base-pairs (bps) 
is central in biological mechanisms~\cite{2005-JPCM-Ambjornsson-Metzler,2006-PRL-Ambjornsson-Metzler,2010-PRL-Jeon-Metzler,
2012-SM-Adamcik-Dietler,2013-PRE-Dasanna-Manghi,2015-JCP-Sicard-Manghi}. 
We performed coarse-grained WT-metaD and Brownian simulations using 
the width $\rho_{\textrm{max}}$ of the bubble defined in Fig.~\ref{fig1}b.
To explore the \textit{slow} entropic contribution associated with the DNA bubble metastable basin we chose to follow the 
evolution of the minimal twist angle $\Phi_{\textrm{min}}$ inside the bubble as auxiliary variable (cf. Fig.~\ref{fig1}b). 
%
The analysis of the FE surface associated with the bubble closure and opening mechanisms, 
as shown in Fig.~\ref{fig1}b,
allowed us to determine the value the FE of formation, $\Delta F = F(op) - F(cl) \approx 9~k_B T$ 
along the MFEP depicted in Fig.~\ref{fig1}b 
and the FE of formation $\Delta F^* \approx 7~k_B T$, defined in term of the probability distribution 
of $\rho_{\textrm{max}}$ and $\Phi_{\textrm{min}}$, 
considering the successive isosurfaces depicted in Fig.~\ref{fig1}b as integration domain. 
The exact values are given in the SM along with the values of 
the parameters $\omega_{op}$ and $\omega_{cl}$ .
Considering the Rouse model~\cite{2005-JPCM-Ambjornsson-Metzler} valid for flexible polymer chain, 
the effective friction coefficient, $\gamma$, in Eq.~\ref{KTconf} depends on the number of opened bps, 
$N_{\textrm{bub}}$, in the DNA bubble. The typical size observed in the simulations, $N_{bub}\approx 10$ bps, 
yields the relation $\gamma_{op}/\gamma_{cl} \approx N_{\textrm{bub}}$ between the effective frictions in Eq.~\ref{KTfinal}.
One obtains the transition rate ratio, $k_{cl \to op}/k_{op \to cl} = (1.5 \pm 0.6) \times 10^{-3}$, in close agreement 
with the numerical ratio obtained within the accelerated MD framework, 
$k^{(num)}_{cl \to op}/k^{(num)}_{op \to cl} = (1.8 \pm 0.4) \times 10^{-3}$ and the experimental times measured 
by Altan-Bonnet \textit{et al.}~\cite{2003-PRL-Altan-Krichevsky}.  
For instance, the \textit{standard} KT yields $k^{(st)}_{\beta \to \alpha}/k^{(st)}_{\alpha \to \beta} = (4.0 \pm 0.7) \times 10^{-3}$, 
which does not differ significantly from our approach. 

\textbf{Circular DNA denaturation bubble.} To conclude this analysis, we studied the cooperative opening and closure 
of denaturation bubble in a negatively supercoiled DNA minicircle within the WT-metaD framework 
and using the width $\rho_{\textrm{max}}$ of the bubble as a CV. As discussed in the SM, we set the parameters 
of the system so that the convergence of the FE profile cannot be reached due to the large configurational entropy 
contribution and the \textit{standard} KT does not apply.
Nevertheless, the convergence of the FE surface in the vicinity of the MFEP shown in Fig.~\ref{fig1}c allowed us to 
to determine the value the FE of formation along the MFEP, $\Delta F = F(op) - F(cl) \approx -4.5~k_B T$ and the FE of formation 
$\Delta F^* \approx -8.5~k_B T$, defined in term of the probability distribution 
of $\rho_{\textrm{max}}$ and $\Phi_{\textrm{min}}$. Considering the typical size of the DNA bubble observed 
in the simulations, $N_{bub}\approx 12$ bps, we determined the parameter $\gamma_{op}/\gamma_{cl}$ 
and obtained the transition rate ratio, $k_{cl \to op}/k_{op \to cl} = (1.0 \pm 0.4) \times 10^{6}$. 
This result is consistent with the \textit{inversion} of the thermodynamic stability of the system 
with respect to opened and closed DNA states, characteristic of the predominant stability of the \textit{long-lived} denaturation 
bubble in supercoiled DNA.
Although the analysis of a converged FE surface was achievable in such case with the appropriate use of the auxiliary variable, 
$\Phi_{\textrm{min}}$, the direct numerical estimation of the transition rates was not achievable with accelerated MD 
approaches, as the shape of the original FE surface could not be evenly maintained after the addition of 
the bias potential~\cite{2010-JCP-Xin-Hamelberg}.\\

In this paper we discussed the theoretical background and algorithmic details to compute the transition rates 
of complex systems when \textit{slow} entropic contribution, such as configurational entropy, comes into play. 
We considered three illustrative applications presenting different level of coarse-graining and entropic contribution.
In the limit where the \textit{slow} entropy contribution does not prevent the reconstruction of a converged FE profile, 
our approach and the \textit{standard} KT showed good agreement with simulations and experiments. 
In the limit of \textit{large} entropic fluctuations, where the shape of the original 
FE landscape cannot be \textit{evenly} maintained within the accelerated MD framework, we showed that our approach could present 
significant improvement over the \textit{standard} KT.
We chose to reconstruct the MFEP and to compute the FE defined in terms of the probability 
distribution of the CVs able to adequately describe the transitions between the FE basins 
and the auxiliary variables used to decipher the \textit{slow} entropic contribution to the system. 
In principle, this would be equivalent to find first the MFEP within the Transition Path sampling framework~\cite{2010-ARPC-Weinan-Vanden} 
and to explore afterwards the entropic properties of the FE landscape~\cite{2017-Elsevier-Peters}.
%
The choice of a specific framework would be motivated by the complexity of the underlying dynamics of the systems. 

Finally, let us comment on the dependence of the measure of the configurational entropy contribution on the choice 
of the auxiliary CVs. 
Similarly to the metaD framework used to explore the FE landscape of complex systems, the reliability of our approach 
is strongly influenced by the choice of the auxiliary CVs considered in Eq.~\ref{Proba-definition}. 
To overcome such limitations, one could consider the potential energy of the system as an auxiliary CV as recently 
explored by Salvalaglio and coworkers~\cite{2018-arXiv-Gimondi-Salvalaglio}, within the metaD framework, 
to break down FE surfaces into their entropic and enthalpic components. Eventually, one would compute rigorously 
the configurational entropy contribution and identify a complementary measure along an arbitrary chosen CV. 
This roadmap will be considered in the near future.\\

The author acknowledges Matteo Salvalaglio for fruitful suggestions and stimulating discussions 
and Nicolas Destainville and Fabio Pietrucci for useful comments. 
Via our membership of the UK's HEC Materials Chemistry Consortium, which is funded by EPSRC (EP/L000202), 
this work used the ARCHER UK National Supercomputing Service (http://www.archer.ac.uk). 

\bibliography{rsc} 

\begin{thebibliography}{42}
\expandafter\ifx\csname natexlab\endcsname\relax\def\natexlab#1{#1}\fi
\expandafter\ifx\csname bibnamefont\endcsname\relax
  \def\bibnamefont#1{#1}\fi
\expandafter\ifx\csname bibfnamefont\endcsname\relax
  \def\bibfnamefont#1{#1}\fi
\expandafter\ifx\csname citenamefont\endcsname\relax
  \def\citenamefont#1{#1}\fi
\expandafter\ifx\csname url\endcsname\relax
  \def\url#1{\texttt{#1}}\fi
\expandafter\ifx\csname urlprefix\endcsname\relax\def\urlprefix{URL }\fi
\providecommand{\bibinfo}[2]{#2}
\providecommand{\eprint}[2][]{\url{#2}}

\bibitem[{\citenamefont{Kramers}(1940)}]{1940-Physica-Kramers}
\bibinfo{author}{\bibfnamefont{H.}~\bibnamefont{Kramers}},
  \bibinfo{journal}{Physica} \textbf{\bibinfo{volume}{7}}, \bibinfo{pages}{284}
  (\bibinfo{year}{1940}).

\bibitem[{\citenamefont{H\"anggi et~al.}(1990)\citenamefont{H\"anggi, Talkner,
  and Borkovec}}]{1990-RMP-Hanggi-Borkovec}
\bibinfo{author}{\bibfnamefont{P.}~\bibnamefont{H\"anggi}},
  \bibinfo{author}{\bibfnamefont{P.}~\bibnamefont{Talkner}}, \bibnamefont{and}
  \bibinfo{author}{\bibfnamefont{M.}~\bibnamefont{Borkovec}},
  \bibinfo{journal}{Rev. Mod. Phys.} \textbf{\bibinfo{volume}{62}},
  \bibinfo{pages}{251} (\bibinfo{year}{1990}).

\bibitem[{\citenamefont{Weinan and
  Vanden-Eijnden}(2010)}]{2010-ARPC-Weinan-Vanden}
\bibinfo{author}{\bibfnamefont{E.}~\bibnamefont{Weinan}} \bibnamefont{and}
  \bibinfo{author}{\bibfnamefont{E.}~\bibnamefont{Vanden-Eijnden}},
  \bibinfo{journal}{Annu. Rev. Phys. Chem.} \textbf{\bibinfo{volume}{61}},
  \bibinfo{pages}{391} (\bibinfo{year}{2010}).

\bibitem[{\citenamefont{Xin et~al.}(2010)\citenamefont{Xin, Doshi, and
  Hamelberg}}]{2010-JCP-Xin-Hamelberg}
\bibinfo{author}{\bibfnamefont{Y.}~\bibnamefont{Xin}},
  \bibinfo{author}{\bibfnamefont{U.}~\bibnamefont{Doshi}}, \bibnamefont{and}
  \bibinfo{author}{\bibfnamefont{D.}~\bibnamefont{Hamelberg}},
  \bibinfo{journal}{J. Chem. Phys.} \textbf{\bibinfo{volume}{132}},
  \bibinfo{pages}{224101} (\bibinfo{year}{2010}).

\bibitem[{\citenamefont{Gobbo et~al.}(2012)\citenamefont{Gobbo, Laio, Maleki,
  and Baroni}}]{2012-PRL-Gobbo-Baroni}
\bibinfo{author}{\bibfnamefont{G.}~\bibnamefont{Gobbo}},
  \bibinfo{author}{\bibfnamefont{A.}~\bibnamefont{Laio}},
  \bibinfo{author}{\bibfnamefont{A.}~\bibnamefont{Maleki}}, \bibnamefont{and}
  \bibinfo{author}{\bibfnamefont{S.}~\bibnamefont{Baroni}},
  \bibinfo{journal}{Phys. Rev. Lett.} \textbf{\bibinfo{volume}{109}},
  \bibinfo{pages}{150601} (\bibinfo{year}{2012}).

\bibitem[{\citenamefont{Tiwary and
  Parrinello}(2013)}]{2013-PRL-Tiwary-Parrinello}
\bibinfo{author}{\bibfnamefont{P.}~\bibnamefont{Tiwary}} \bibnamefont{and}
  \bibinfo{author}{\bibfnamefont{M.}~\bibnamefont{Parrinello}},
  \bibinfo{journal}{Phys. Rev. Lett.} \textbf{\bibinfo{volume}{111}},
  \bibinfo{pages}{230602} (\bibinfo{year}{2013}).

\bibitem[{\citenamefont{Salvalaglio et~al.}(2014)\citenamefont{Salvalaglio,
  Tiwary, and Parrinello}}]{2014-JCTC-Salvalaglio-Parrinello}
\bibinfo{author}{\bibfnamefont{M.}~\bibnamefont{Salvalaglio}},
  \bibinfo{author}{\bibfnamefont{P.}~\bibnamefont{Tiwary}}, \bibnamefont{and}
  \bibinfo{author}{\bibfnamefont{M.}~\bibnamefont{Parrinello}},
  \bibinfo{journal}{J. Chem. Theory Comput.} \textbf{\bibinfo{volume}{10}},
  \bibinfo{pages}{1420} (\bibinfo{year}{2014}).

\bibitem[{\citenamefont{Sicard et~al.}(2015)\citenamefont{Sicard, Destainville,
  and Manghi}}]{2015-JCP-Sicard-Manghi}
\bibinfo{author}{\bibfnamefont{F.}~\bibnamefont{Sicard}},
  \bibinfo{author}{\bibfnamefont{N.}~\bibnamefont{Destainville}},
  \bibnamefont{and} \bibinfo{author}{\bibfnamefont{M.}~\bibnamefont{Manghi}},
  \bibinfo{journal}{J. Chem. Phys.} \textbf{\bibinfo{volume}{142}},
  \bibinfo{pages}{034903} (\bibinfo{year}{2015}).

\bibitem[{\citenamefont{M\"okk\"onen et~al.}(2016)\citenamefont{M\"okk\"onen,
  Ala-Nissila, and J\'onsson}}]{2016-JCP-Mokkonen-Jonsson}
\bibinfo{author}{\bibfnamefont{H.}~\bibnamefont{M\"okk\"onen}},
  \bibinfo{author}{\bibfnamefont{T.}~\bibnamefont{Ala-Nissila}},
  \bibnamefont{and}
  \bibinfo{author}{\bibfnamefont{H.}~\bibnamefont{J\'onsson}},
  \bibinfo{journal}{J. Chem. Phys.} \textbf{\bibinfo{volume}{145}},
  \bibinfo{pages}{094901} (\bibinfo{year}{2016}).

\bibitem[{\citenamefont{Pietrucci}(2017)}]{2017-RP-Pietrucci}
\bibinfo{author}{\bibfnamefont{F.}~\bibnamefont{Pietrucci}},
  \bibinfo{journal}{Rev. Phys.} \textbf{\bibinfo{volume}{2}},
  \bibinfo{pages}{32} (\bibinfo{year}{2017}).

\bibitem[{\citenamefont{Truhlar et~al.}(1996)\citenamefont{Truhlar, Garrett,
  and Klippenstein}}]{1996-JPC-Truhlar-Klippenstein}
\bibinfo{author}{\bibfnamefont{D.}~\bibnamefont{Truhlar}},
  \bibinfo{author}{\bibfnamefont{B.}~\bibnamefont{Garrett}}, \bibnamefont{and}
  \bibinfo{author}{\bibfnamefont{S.}~\bibnamefont{Klippenstein}},
  \bibinfo{journal}{J. Phys. Chem.} \textbf{\bibinfo{volume}{100}},
  \bibinfo{pages}{12771} (\bibinfo{year}{1996}).

\bibitem[{\citenamefont{Peters}(2017)}]{2017-Elsevier-Peters}
\bibinfo{author}{\bibfnamefont{B.}~\bibnamefont{Peters}},
  \emph{\bibinfo{title}{Reaction Rate Theory and Rare Events, 1st Ed.}}
  (\bibinfo{publisher}{Elsevier: Amsterdam, The Netherlands},
  \bibinfo{year}{2017}).

\bibitem[{\citenamefont{De et~al.}(2017)\citenamefont{De, Singh, and
  Gupta}}]{2017-arXiv-De-Gupta}
\bibinfo{author}{\bibfnamefont{D.}~\bibnamefont{De}},
  \bibinfo{author}{\bibfnamefont{A.}~\bibnamefont{Singh}}, \bibnamefont{and}
  \bibinfo{author}{\bibfnamefont{A.}~\bibnamefont{Gupta}},
  \bibinfo{journal}{arXiv:1705.01246}  (\bibinfo{year}{2017}).

\bibitem[{\citenamefont{Karplus and
  Kushick}(1981)}]{1981-Macromolecules-Karplus-Kushick}
\bibinfo{author}{\bibfnamefont{M.}~\bibnamefont{Karplus}} \bibnamefont{and}
  \bibinfo{author}{\bibfnamefont{J.}~\bibnamefont{Kushick}},
  \bibinfo{journal}{Macromolecules} \textbf{\bibinfo{volume}{14}},
  \bibinfo{pages}{325} (\bibinfo{year}{1981}).

\bibitem[{\citenamefont{Nguyen and
  Derreumaux}(2012)}]{2012-PCCP-Nguyen-Derreumaux}
\bibinfo{author}{\bibfnamefont{P.}~\bibnamefont{Nguyen}} \bibnamefont{and}
  \bibinfo{author}{\bibfnamefont{P.}~\bibnamefont{Derreumaux}},
  \bibinfo{journal}{Phys. Chem. Chem. Phys.} \textbf{\bibinfo{volume}{14}},
  \bibinfo{pages}{877} (\bibinfo{year}{2012}).

\bibitem[{\citenamefont{Laio and Parrinello}(2002)}]{2002-PNAS-Laio-Parrinello}
\bibinfo{author}{\bibfnamefont{A.}~\bibnamefont{Laio}} \bibnamefont{and}
  \bibinfo{author}{\bibfnamefont{M.}~\bibnamefont{Parrinello}},
  \bibinfo{journal}{Proc. Nat. Acad. Soc. U.S.A.}
  \textbf{\bibinfo{volume}{99}}, \bibinfo{pages}{12562} (\bibinfo{year}{2002}).

\bibitem[{\citenamefont{Laio and Gervasio}(2008)}]{2008-RPP-Laio-Gervasio}
\bibinfo{author}{\bibfnamefont{A.}~\bibnamefont{Laio}} \bibnamefont{and}
  \bibinfo{author}{\bibfnamefont{F.}~\bibnamefont{Gervasio}},
  \bibinfo{journal}{Rep. Prog. Phys.} \textbf{\bibinfo{volume}{71}},
  \bibinfo{pages}{126601} (\bibinfo{year}{2008}).

\bibitem[{\citenamefont{Gimondi and
  Salvalaglio}(2017)}]{2017-JCP-Gimondi-Salvalaglio}
\bibinfo{author}{\bibfnamefont{I.}~\bibnamefont{Gimondi}} \bibnamefont{and}
  \bibinfo{author}{\bibfnamefont{M.}~\bibnamefont{Salvalaglio}},
  \bibinfo{journal}{J. Chem. Phys.} \textbf{\bibinfo{volume}{147}},
  \bibinfo{pages}{114502} (\bibinfo{year}{2017}).

\bibitem[{\citenamefont{Coffey and Kalmykov}(2012)}]{2012-WS-Coffey-Kalmykov}
\bibinfo{author}{\bibfnamefont{W.}~\bibnamefont{Coffey}} \bibnamefont{and}
  \bibinfo{author}{\bibfnamefont{Y.}~\bibnamefont{Kalmykov}},
  \emph{\bibinfo{title}{The Langevin Equation: With Applications to Stochastic
  Problems in Physics, Chemistry and Electrical Engineering, 3rd Ed.; World
  Scientific Series in Contemporary Chemical Physics}},
  vol.~\bibinfo{volume}{27} (\bibinfo{publisher}{World Scientific Publishing
  Company: Singapore}, \bibinfo{year}{2012}).

\bibitem[{\citenamefont{Marini et~al.}(2008)\citenamefont{Marini, Marconi,
  Rondoni, and Vulpiani}}]{2008-PR-Marini-Vulpiani}
\bibinfo{author}{\bibfnamefont{U.}~\bibnamefont{Marini}},
  \bibinfo{author}{\bibfnamefont{B.}~\bibnamefont{Marconi}},
  \bibinfo{author}{\bibfnamefont{A.~P.~L.} \bibnamefont{Rondoni}},
  \bibnamefont{and} \bibinfo{author}{\bibfnamefont{A.}~\bibnamefont{Vulpiani}},
  \bibinfo{journal}{Phys. Rep.} \textbf{\bibinfo{volume}{461}},
  \bibinfo{pages}{111} (\bibinfo{year}{2008}).

\bibitem[{\citenamefont{Woodside et~al.}(2014)\citenamefont{Woodside, Lambert,
  and Beach}}]{2014-BJ-Woodside-Beach}
\bibinfo{author}{\bibfnamefont{M.}~\bibnamefont{Woodside}},
  \bibinfo{author}{\bibfnamefont{J.}~\bibnamefont{Lambert}}, \bibnamefont{and}
  \bibinfo{author}{\bibfnamefont{K.}~\bibnamefont{Beach}},
  \bibinfo{journal}{Biophys. J.} \textbf{\bibinfo{volume}{107}},
  \bibinfo{pages}{1647} (\bibinfo{year}{2014}).

\bibitem[{\citenamefont{Flaherty and
  Iglesia}(2013)}]{2013-JACS-Flaherty-Iglesia}
\bibinfo{author}{\bibfnamefont{D.}~\bibnamefont{Flaherty}} \bibnamefont{and}
  \bibinfo{author}{\bibfnamefont{E.}~\bibnamefont{Iglesia}},
  \bibinfo{journal}{J. Am. Chem. Soc.} \textbf{\bibinfo{volume}{135}},
  \bibinfo{pages}{18586} (\bibinfo{year}{2013}).

\bibitem[{\citenamefont{Aqvist et~al.}(2017)\citenamefont{Aqvist, Kazemi,
  Isaken, and Brandsdal}}]{2017-ACR-Aqvist-Brandsdal}
\bibinfo{author}{\bibfnamefont{J.}~\bibnamefont{Aqvist}},
  \bibinfo{author}{\bibfnamefont{M.}~\bibnamefont{Kazemi}},
  \bibinfo{author}{\bibfnamefont{G.}~\bibnamefont{Isaken}}, \bibnamefont{and}
  \bibinfo{author}{\bibfnamefont{B.}~\bibnamefont{Brandsdal}},
  \bibinfo{journal}{Acc. Chem. Res.} \textbf{\bibinfo{volume}{50}},
  \bibinfo{pages}{199} (\bibinfo{year}{2017}).

\bibitem[{\citenamefont{Sensale et~al.}(2017)\citenamefont{Sensale, Peng, and
  Chang}}]{2017-JCP-Sensale-Chang}
\bibinfo{author}{\bibfnamefont{S.}~\bibnamefont{Sensale}},
  \bibinfo{author}{\bibfnamefont{Z.}~\bibnamefont{Peng}}, \bibnamefont{and}
  \bibinfo{author}{\bibfnamefont{H.-C.} \bibnamefont{Chang}},
  \bibinfo{journal}{J. Chem. Phys.} \textbf{\bibinfo{volume}{147}},
  \bibinfo{pages}{135101} (\bibinfo{year}{2017}).

\bibitem[{\citenamefont{Leuzzi and
  Nieuwenhuizen}(2007)}]{2007-CRC-Leuzzi-Nieuwenhuizen}
\bibinfo{author}{\bibfnamefont{L.}~\bibnamefont{Leuzzi}} \bibnamefont{and}
  \bibinfo{author}{\bibfnamefont{T.}~\bibnamefont{Nieuwenhuizen}},
  \emph{\bibinfo{title}{Thermodynamics of the Glassy State, 1st Ed.; World
  Scientific Series in Contemporary Chemical Physics}}
  (\bibinfo{publisher}{Taylor and Francis Group, CRC Press: Boca Raton (FL)},
  \bibinfo{year}{2007}).

\bibitem[{\citenamefont{Peter et~al.}(2004)\citenamefont{Peter, Oostenbrink,
  van Dorp, and van Gunsteren}}]{2004-JCP-Peter-vanGunsteren}
\bibinfo{author}{\bibfnamefont{C.}~\bibnamefont{Peter}},
  \bibinfo{author}{\bibfnamefont{C.}~\bibnamefont{Oostenbrink}},
  \bibinfo{author}{\bibfnamefont{A.}~\bibnamefont{van Dorp}}, \bibnamefont{and}
  \bibinfo{author}{\bibfnamefont{W.}~\bibnamefont{van Gunsteren}},
  \bibinfo{journal}{J. Chem. Phys.} \textbf{\bibinfo{volume}{120}},
  \bibinfo{pages}{2652} (\bibinfo{year}{2004}).

\bibitem[{\citenamefont{Cheluvaraja and
  Meirovitch}(2006)}]{2006-JCP-Cheluvaraja-Meirovitch}
\bibinfo{author}{\bibfnamefont{S.}~\bibnamefont{Cheluvaraja}} \bibnamefont{and}
  \bibinfo{author}{\bibfnamefont{H.}~\bibnamefont{Meirovitch}},
  \bibinfo{journal}{J. Chem. Phys.} \textbf{\bibinfo{volume}{122}},
  \bibinfo{pages}{054903} (\bibinfo{year}{2006}).

\bibitem[{\citenamefont{Bolhuis et~al.}(2000)\citenamefont{Bolhuis, Dellago,
  and Chandler}}]{2000-PNAS-Bolhuis-Chandler}
\bibinfo{author}{\bibfnamefont{P.}~\bibnamefont{Bolhuis}},
  \bibinfo{author}{\bibfnamefont{C.}~\bibnamefont{Dellago}}, \bibnamefont{and}
  \bibinfo{author}{\bibfnamefont{D.}~\bibnamefont{Chandler}},
  \bibinfo{journal}{Proc. Nat. Acad. Sci. U.S.A} \textbf{\bibinfo{volume}{97}},
  \bibinfo{pages}{5877} (\bibinfo{year}{2000}).

\bibitem[{\citenamefont{Chen et~al.}(2013)\citenamefont{Chen, Huang, and
  Xiao}}]{2013-JCP-Chen-Xiao}
\bibinfo{author}{\bibfnamefont{C.}~\bibnamefont{Chen}},
  \bibinfo{author}{\bibfnamefont{Y.}~\bibnamefont{Huang}}, \bibnamefont{and}
  \bibinfo{author}{\bibfnamefont{Y.}~\bibnamefont{Xiao}}, \bibinfo{journal}{J.
  Chem. Phys.} \textbf{\bibinfo{volume}{138}}, \bibinfo{pages}{164122}
  (\bibinfo{year}{2013}).

\bibitem[{\citenamefont{Landau and Lifshitz}(1980)}]{1980-Landau-Lifshitz}
\bibinfo{author}{\bibfnamefont{L.}~\bibnamefont{Landau}} \bibnamefont{and}
  \bibinfo{author}{\bibfnamefont{E.}~\bibnamefont{Lifshitz}},
  \emph{\bibinfo{title}{Statistical Physics, 3rd Ed., Revised and Enlarged}}
  (\bibinfo{publisher}{Butterworth-Heinemann, Oxford (UK)},
  \bibinfo{year}{1980}).

\bibitem[{\citenamefont{Laldler and King}(1983)}]{1983-JPC-Laldler-King}
\bibinfo{author}{\bibfnamefont{K.}~\bibnamefont{Laldler}} \bibnamefont{and}
  \bibinfo{author}{\bibfnamefont{M.}~\bibnamefont{King}}, \bibinfo{journal}{J.
  Phys. Chem.} \textbf{\bibinfo{volume}{87}}, \bibinfo{pages}{2657}
  (\bibinfo{year}{1983}).

\bibitem[{\citenamefont{Ren et~al.}(2005)\citenamefont{Ren, Vanden-Eijnden,
  Maragakis, and Weinan}}]{2005-JCP-Ren-Weinan}
\bibinfo{author}{\bibfnamefont{W.}~\bibnamefont{Ren}},
  \bibinfo{author}{\bibfnamefont{E.}~\bibnamefont{Vanden-Eijnden}},
  \bibinfo{author}{\bibfnamefont{P.}~\bibnamefont{Maragakis}},
  \bibnamefont{and} \bibinfo{author}{\bibfnamefont{E.}~\bibnamefont{Weinan}},
  \bibinfo{journal}{J. Chem. Phys.} \textbf{\bibinfo{volume}{123}},
  \bibinfo{pages}{134109} (\bibinfo{year}{2005}).

\bibitem[{\citenamefont{Cuny et~al.}(2017)\citenamefont{Cuny, Korchagina,
  Menakbi, and Mineva}}]{2017-JMM-Cuny-Mineva}
\bibinfo{author}{\bibfnamefont{J.}~\bibnamefont{Cuny}},
  \bibinfo{author}{\bibfnamefont{K.}~\bibnamefont{Korchagina}},
  \bibinfo{author}{\bibfnamefont{C.}~\bibnamefont{Menakbi}}, \bibnamefont{and}
  \bibinfo{author}{\bibfnamefont{T.}~\bibnamefont{Mineva}},
  \bibinfo{journal}{J. Mol. Model} \textbf{\bibinfo{volume}{23}},
  \bibinfo{pages}{72} (\bibinfo{year}{2017}).

\bibitem[{\citenamefont{Gimondi et~al.}(2018)\citenamefont{Gimondi, Tribello,
  and Salvalaglio}}]{2018-arXiv-Gimondi-Salvalaglio}
\bibinfo{author}{\bibfnamefont{I.}~\bibnamefont{Gimondi}},
  \bibinfo{author}{\bibfnamefont{G.}~\bibnamefont{Tribello}}, \bibnamefont{and}
  \bibinfo{author}{\bibfnamefont{M.}~\bibnamefont{Salvalaglio}},
  \bibinfo{journal}{arXiv:1803.01093 [cond-mat.stat-mech]}
  (\bibinfo{year}{2018}).

\bibitem[{\citenamefont{Barducci et~al.}(2008)\citenamefont{Barducci, Bussi,
  and Parrinello}}]{2008-PRL-Barducci-Parrinello}
\bibinfo{author}{\bibfnamefont{A.}~\bibnamefont{Barducci}},
  \bibinfo{author}{\bibfnamefont{G.}~\bibnamefont{Bussi}}, \bibnamefont{and}
  \bibinfo{author}{\bibfnamefont{M.}~\bibnamefont{Parrinello}},
  \bibinfo{journal}{Phys. Rev. Lett.} \textbf{\bibinfo{volume}{100}},
  \bibinfo{pages}{020603} (\bibinfo{year}{2008}).

\bibitem[{\citenamefont{Dama et~al.}(2014)\citenamefont{Dama, Parrinello, and
  Voth}}]{2014-PRL-Dama-Voth}
\bibinfo{author}{\bibfnamefont{J.}~\bibnamefont{Dama}},
  \bibinfo{author}{\bibfnamefont{M.}~\bibnamefont{Parrinello}},
  \bibnamefont{and} \bibinfo{author}{\bibfnamefont{G.}~\bibnamefont{Voth}},
  \bibinfo{journal}{Phys. Rev. Lett.} \textbf{\bibinfo{volume}{112}},
  \bibinfo{pages}{240602} (\bibinfo{year}{2014}).

\bibitem[{\citenamefont{Ambjornssson and
  Metzler}(2005)}]{2005-JPCM-Ambjornsson-Metzler}
\bibinfo{author}{\bibfnamefont{T.}~\bibnamefont{Ambjornssson}}
  \bibnamefont{and} \bibinfo{author}{\bibfnamefont{R.}~\bibnamefont{Metzler}},
  \bibinfo{journal}{J. Phys. Condens. Matter} \textbf{\bibinfo{volume}{17}},
  \bibinfo{pages}{S1841} (\bibinfo{year}{2005}).

\bibitem[{\citenamefont{Ambjornsson et~al.}(2006)\citenamefont{Ambjornsson,
  Banik, Krichevsky, and Metzler}}]{2006-PRL-Ambjornsson-Metzler}
\bibinfo{author}{\bibfnamefont{T.}~\bibnamefont{Ambjornsson}},
  \bibinfo{author}{\bibfnamefont{S.}~\bibnamefont{Banik}},
  \bibinfo{author}{\bibfnamefont{O.}~\bibnamefont{Krichevsky}},
  \bibnamefont{and} \bibinfo{author}{\bibfnamefont{R.}~\bibnamefont{Metzler}},
  \bibinfo{journal}{Phys. Rev. Lett.} \textbf{\bibinfo{volume}{97}},
  \bibinfo{pages}{128105} (\bibinfo{year}{2006}).

\bibitem[{\citenamefont{Jeon et~al.}(2010)\citenamefont{Jeon, Adamcik, Dietler,
  and Metzler}}]{2010-PRL-Jeon-Metzler}
\bibinfo{author}{\bibfnamefont{J.-H.} \bibnamefont{Jeon}},
  \bibinfo{author}{\bibfnamefont{J.}~\bibnamefont{Adamcik}},
  \bibinfo{author}{\bibfnamefont{G.}~\bibnamefont{Dietler}}, \bibnamefont{and}
  \bibinfo{author}{\bibfnamefont{R.}~\bibnamefont{Metzler}},
  \bibinfo{journal}{Phys. Rev. Lett.} \textbf{\bibinfo{volume}{105}},
  \bibinfo{pages}{208101} (\bibinfo{year}{2010}).

\bibitem[{\citenamefont{Adamcik et~al.}(2012)\citenamefont{Adamcik, Jeon,
  Karczewski, Metzler, and Dietler}}]{2012-SM-Adamcik-Dietler}
\bibinfo{author}{\bibfnamefont{J.}~\bibnamefont{Adamcik}},
  \bibinfo{author}{\bibfnamefont{J.-H.} \bibnamefont{Jeon}},
  \bibinfo{author}{\bibfnamefont{K.}~\bibnamefont{Karczewski}},
  \bibinfo{author}{\bibfnamefont{R.}~\bibnamefont{Metzler}}, \bibnamefont{and}
  \bibinfo{author}{\bibfnamefont{G.}~\bibnamefont{Dietler}},
  \bibinfo{journal}{Soft Matter} \textbf{\bibinfo{volume}{8}},
  \bibinfo{pages}{8651} (\bibinfo{year}{2012}).

\bibitem[{\citenamefont{Dasanna et~al.}(2013)\citenamefont{Dasanna,
  Destainville, Palmeri, and Manghi}}]{2013-PRE-Dasanna-Manghi}
\bibinfo{author}{\bibfnamefont{A.}~\bibnamefont{Dasanna}},
  \bibinfo{author}{\bibfnamefont{N.}~\bibnamefont{Destainville}},
  \bibinfo{author}{\bibfnamefont{J.}~\bibnamefont{Palmeri}}, \bibnamefont{and}
  \bibinfo{author}{\bibfnamefont{M.}~\bibnamefont{Manghi}},
  \bibinfo{journal}{Phys. Rev. E} \textbf{\bibinfo{volume}{87}},
  \bibinfo{pages}{052703} (\bibinfo{year}{2013}).

\bibitem[{\citenamefont{Altan-Bonnet et~al.}(2003)\citenamefont{Altan-Bonnet,
  Libchaber, and Krichevsky}}]{2003-PRL-Altan-Krichevsky}
\bibinfo{author}{\bibfnamefont{G.}~\bibnamefont{Altan-Bonnet}},
  \bibinfo{author}{\bibfnamefont{A.}~\bibnamefont{Libchaber}},
  \bibnamefont{and}
  \bibinfo{author}{\bibfnamefont{O.}~\bibnamefont{Krichevsky}},
  \bibinfo{journal}{Phys. Rev. Lett.} \textbf{\bibinfo{volume}{90}},
  \bibinfo{pages}{138101} (\bibinfo{year}{2003}).

\end{thebibliography}


\begin{thebibliography}{99}
%
\bibitem{SM-2005-JCP-Ren-Weinan} W. Ren, E. Vanden-Eijnden, P. Maragakis and E. Weinan, J. Chem. Phys. \textbf{123}, 134109 (2005).
\bibitem{SM-2013-PRL-Tiwary-Parrinello} P. Tiwary and M. Parrinello, Phys. Rev. Lett. \textbf{111}, 230602 (2013).
\bibitem{SM-2014-JCTC-Salvalaglio-Parrinello} M. Salvalaglio, P. Tiwary and M. Parrinello, J. Chem. Theory Comput. \textbf{10}, 1420-1425 (2014).
\bibitem{SM-2017-JMM-Cuny-Mineva} J. Cuny, K. Korchagina, C. Menakbi and T. Mineva, J. Mol. Model \textbf{23}, 72 (2017).
\bibitem{SM-2018-arXiv-Gimondi-Salvalaglio} I. Gimondi, G.A. Tribello and M. Salvalaglio, arXiv:1803.01093 [cond-mat.stat-mech] (2018).
\bibitem{SM-2007-JCP-Bussi-Parrinello} G. Bussi, D. Donadio, and M. Parrinello, J. Chem. Phys. \textbf{126}, 014101 (2007).
\bibitem{SM-2005-JCC-Case-Woods} D.A. Case et al., J. Comp. Chem. \textbf{26}, 1668-1688 (2005).
\bibitem{SM-2001-JMM-Lindahl-VanDerSpoel} E. Lindahl, B. Hess, and D. Van Der Spoel, J. Mol. Model. \textbf{7},306-317.
\bibitem{SM-2014-CPC-Tribello-Bussi} G.A. Tribello, M. Bonomi, D. Branduardi, C. Camilloni and G. Bussi, Comput. Phys. Comm. \textbf{185}, 604-613 (2014).
\bibitem{SM-2008-PRL-Barducci-Parrinello} A. Barducci, G. Bussi, and M. Parrinello, Phys. Rev. Lett. \textbf{100}, 020603 (2008).
\bibitem{SM-2014-PRL-Dama-Voth} J.F. Dama, M. Parrinello, and G.A. Voth, Phys. Rev. Lett. \textbf{112}, 240602 (2014).
\bibitem{SM-1997-JCC-Hess-Fraaije} B. Hess, H. Bekker, H.J. Berendsen, and J.G. Fraaije, J. Comput. Chem. \textbf{98}, 1463-1472 (1997).
\bibitem{SM-1993-JCP-Darden-Pedersen} T. Darden, D. York, and L. Pedersen, J. Chem. Phys. \textbf{135}, 145102 (1993).
\bibitem{SM-2013-JCP-Chen-Xiao} C. Chen, Y. Huang, and Y. Xiao, J. Chem. Phys. \textbf{138}, 164122 (2013).
\bibitem{SM-2010-JCP-Xin-Hamelberg} Y. Xin, U. Doshi, and D. Hamelberg, J. Chem. Phys. \textbf{132}, 224101 (2010).
\bibitem{SM-2000-Bolhuis-Chandler} P.G. Bolhuis, C. Dellago, and D. Chandler, Proc. Nat. Acad. Sci. \textbf{97}, 5877-5882 (2000).
\bibitem{SM-2003-PRL-Altan-Krichevsky} G. Altan-Bonnet,  A. Libchaber, and O. Krichevsky, Phys. Rev. Lett.\textbf{90}, 138101 (2003).
\bibitem{SM-2013-PRE-Dasanna-Manghi} A.K. Dasanna, N. Destainville, and J. Palmeri and M. Manghi, Phys. Rev. E \textbf{87}, 052703 (2013).
\bibitem{SM-2015-JCP-Sicard-Manghi} F. Sicard, N. Destainville and M. Manghi, J. Chem. Phys. \textbf{142}, 034903 (2015).
\bibitem{SM-Hugel-PRL2005} T. Hugel, M. Rief, M. Seitz, H. E. Gaub, and R. Netz, Phys. Rev. Lett. \textbf{94}, 048301 (2005).
\bibitem{SM-Grosberg-AIP1994} A. Y. Grosberg and A. R. Khokhlov, Statistical Physics of Macromolecules (AIP, Melville, NY, 1994).
\bibitem{SM-Dasanna-EPL2012} A. K. Dasanna, N. Destainville, J. Palmeri, and M. Manghi, EuroPhys. Lett. \textbf{98}, 38002 (2012).
\bibitem{SM-Tinland-Macro1997} B. Tinland, A. Pluen, J. Sturm, and G. Weill, Macromolecules \textbf{30}, 5763 (1997).
\bibitem{SM-Murtola-PCCP2009} T. Murtola, A. Bunker, I. Vattulainen, M. Deserno, and M. Karttunen, Phys. Chem. Chem. Phys. \textbf{11}, 1869 (2009).
\bibitem{SM-Bustamante-COSB2000} C. Bustamante, S. B. Smith, J. Liphardt, and D. Smith, Curr. Opin. Struct. Biol. \textbf{10}, 279 (2000).
\bibitem{SM-2009-JCC-Bonomi-Parrinello} M. Bonomi, A. Barducci and M. Parrinello, J Comput. Chem. \textbf{30}, 1615 (2009).
\bibitem{SM-2005-JPCM-Ambjornsson-Metzler} T. Ambjornssson and R. Metzler, J. Phys. Condens. Matter \textbf{17}, S1841 (2005).
%
\end{thebibliography}

\pagebreak
\widetext
\begin{center}
\textbf{\large Computing Transition Rates for Rare Event: When Kramers Theory meets Free Energy Landscape\vskip 0.5cm
			   Supplemental Material}
\end{center}
\setcounter{equation}{0}
\setcounter{figure}{0}
\setcounter{table}{0}
\setcounter{page}{1}
\makeatletter
\renewcommand{\theequation}{S\arabic{equation}}
\renewcommand{\thefigure}{S\arabic{figure}}
\renewcommand{\bibnumfmt}[1]{[S#1]}
\renewcommand{\citenumfont}[1]{S#1}
%
\section{Alanine dipeptide in vacuum}
\begin{figure}[b]
\includegraphics[width=1.0 \textwidth, angle=-0]{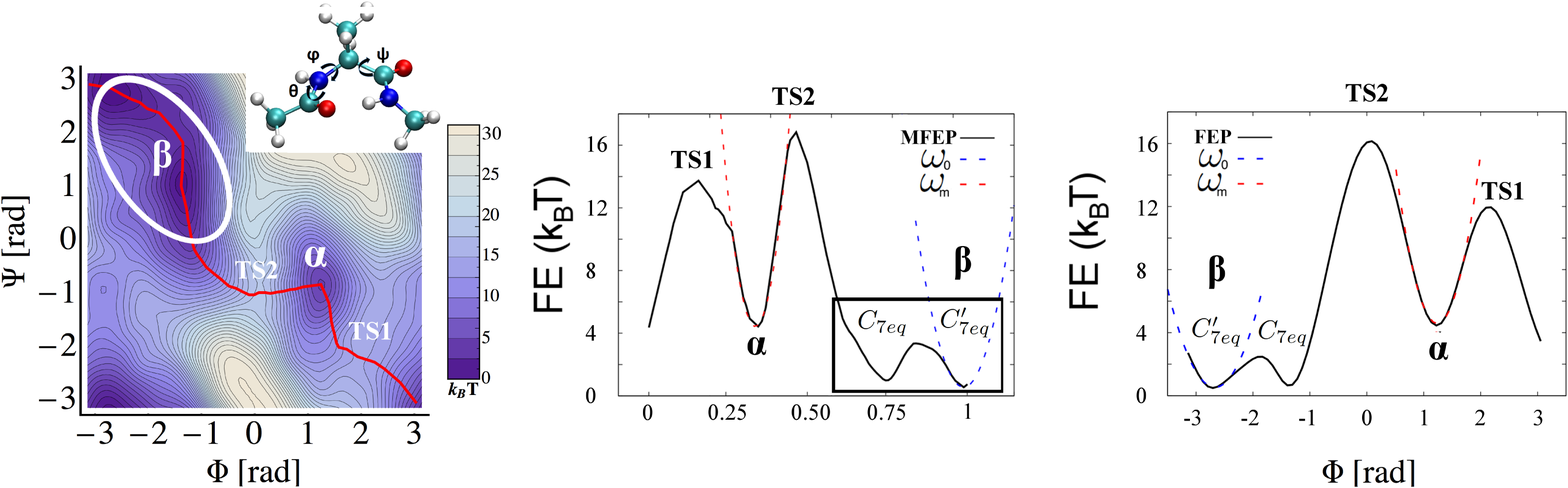}
 \caption{\textbf{Left panel:} FE surface associated with the conformational transition between conformers 
 $\alpha$ and $\beta$ of alanine dipeptide in vacuum as a function of the two dihedral angles $\Phi$ and $\Psi$ (see inset). 
 The contour lines are every half $k_B T$. The typical MFEP obtained within the steepest descent 
 framework~\cite{SM-2013-JCP-Chen-Xiao} is shown in red color along with the locations of the transition states, TS1 and TS2.
 \textbf{Middle panel:} FE profile of the alanine dipeptide in vacuum as a function of the progression along the typical MFEP 
 (normalized to unity) obtained within the steepest descent framework~\cite{SM-2013-JCP-Chen-Xiao}. 
 The nonlinear least-squares Marquardt-Levenberg algorithm was implemented to fit the parameters  $\omega_0$ and $\omega_m$, 
 measured in the equilibrium and metastable states, respectively.
 \textbf{Right panel:} FE profile of alanine dipeptide in vacuum reconstructed along the dihedral angle $\Phi$ obtained within 
the WT-metaD framework. The nonlinear least-squares Marquardt-Levenberg algorithm was implemented to fit the parameters  
$\omega_0$ and $\omega_m$, measured in the equilibrium and metastable states, respectively.
 }
\label{figSM1}
\end{figure}

The conformational transition between conformers $\alpha$ and $\beta$ of the alanine dipeptide molecule 
has been extensively studied as an example of rare event~\cite{SM-2005-JCP-Ren-Weinan, SM-2013-PRL-Tiwary-Parrinello,
SM-2014-JCTC-Salvalaglio-Parrinello,SM-2017-JMM-Cuny-Mineva,SM-2018-arXiv-Gimondi-Salvalaglio}. 
The two stable states are differentiated by the values 
of the backbone dihedral angles $\Phi$ and $\Psi$, as defined in the inset of Fig.~\ref{figSM1} (left panel), 
and are separated by a activation free energy (FE) barrier of $\approx 8$ kcal/mol.
We used a Langevin thermostat to enforce the temperature~\cite{SM-2007-JCP-Bussi-Parrinello}, a time step of $0.2$ fs, 
AMBER03 forcefield~\cite{SM-2005-JCC-Case-Woods} and GROMACS~5.1 molecular dynamics code~\cite{SM-2001-JMM-Lindahl-VanDerSpoel} 
patched with PLUMED~2.3~\cite{SM-2014-CPC-Tribello-Bussi}.
To reconstruct the FE surface, we performed well-tempered metaD (WT-metaD) atomistic 
simulations~\cite{SM-2008-PRL-Barducci-Parrinello,SM-2014-PRL-Dama-Voth} using both torsional angles $\Phi$ and $\Psi$ 
as collective variables (CVs), a bias factor of $15$ at $300$ K. 
The initial Gaussian height was $1.25$ kJ/mol, the width was $0.25$ rad, and the deposition stride was $0.12$ ps.
A single alanine dipeptide molecule was kept in a periodic cubic box of edge $\approx 3$ nm. 
The LINCS algorithm~\cite{SM-1997-JCC-Hess-Fraaije} handled bond constraints while 
the particle-mesh Ewald scheme~\cite{SM-1993-JCP-Darden-Pedersen} was used to treat long-range 
electrostatic interactions. The non-bonded van der Waals cutoff radius was $0.8$ nm.
\\

Fig.~\ref{figSM1} (left panel) shows the FE surface for this molecule, along with the rough locations of the stable states. 
The two minima $C_{7eq}$ and $C'_{7eq}$ are combined in the $\beta$ basin 
as in Refs.~\cite{SM-2000-Bolhuis-Chandler,SM-2013-PRL-Tiwary-Parrinello}.
The location of the metastable basins and the heigh of the FE barrier are in agreement with the ones found 
in the literature~\cite{SM-2005-JCP-Ren-Weinan,SM-2013-PRL-Tiwary-Parrinello,SM-2014-JCTC-Salvalaglio-Parrinello}. 
We determined the value of the FE of formation, $\Delta F^0_{\alpha \beta} = F(\beta) - F(\alpha) = 3.6 \pm 0.4~k_B T$ 
and the activation energies, $\Delta F_{\alpha \to \beta} =  9.1 \pm 0.1~k_B T$ 
and $\Delta F_{\beta \to \alpha} = 12.6 \pm 0.1~k_B T$ along the MFEP obtained within the steepest descent 
framework~\cite{SM-2013-JCP-Chen-Xiao}, as shown in Fig.~\ref{figSM1} (middle panel). The FE of formation, 
$\Delta F^*_{\alpha \beta} = 4.7 \pm 0.1~k_B T$, defined in term of the probability distribution of $\Phi$ and $\Psi$, 
was computed considering the successive isosurfaces in the FE basins depicted in Fig.~\ref{figSM1} (left panel) 
as integration domains (cf. Eq.~(13) in the main text). 
In Fig.~\ref{figSM1} (middle panel), we show the FE of the peptide as a function of the progression along 
the typical MFEP (normalized to unity). The nonlinear least-squares Marquardt-Levenberg algorithm was implemented 
to fit the parameters $\omega_0$ and $\omega_{m}$ with Gaussian distribution. We obtained  
$\omega_{m} = 5.0 \pm 0.1$ and $\omega_0 = 3.3 \pm 0.1$ for the metastable ($\alpha$) and equilibrium ($\beta$) basins, respectively. 
Assuming that the effective friction coefficient, $\gamma$, 
remains unchanged in the transitions $\alpha \leftrightarrow \beta$, one obtains the transition rate ratio, 
$k_{\beta \to \alpha}/k_{\alpha \to \beta} = (5.6 \pm 2.0) \times 10^{-2}$.

In Fig.~\ref{figSM1} (right panel), we show the FE profile of the peptide along the dihedral angle $\Phi$ reconstructed within 
the WT-metaD framework. We determined the value of the FE of formation, $\Delta F^0_{\alpha \beta} = 3.9 \pm 0.1~k_B T$ 
and the activation energies, $\Delta F_{\alpha \to \beta} = 7.7 \pm 0.1~k_B T$ 
and $\Delta F_{\beta \to \alpha} = 11.6 \pm 0.1~k_B T$. The nonlinear least-squares Marquardt-Levenberg algorithm was implemented 
to fit the parameters $\omega_0$ and $\omega_{m}$ with Gaussian distribution. We obtained  
$\omega_{m} = 6.8 \pm 0.1$ and $\omega_0 = 4.7 \pm 0.2$ for the metastable ($\alpha$) and equilibrium ($\beta$) basins, respectively.
The \textit{standard} KT yields $k^{(st)}_{\beta \to \alpha}/k^{(st)}_{\alpha \to \beta} = (1.4 \pm 0.2) \times 10^{-2}$.

We extended the Metadynamics scope~\cite{SM-2010-JCP-Xin-Hamelberg,SM-2013-PRL-Tiwary-Parrinello,SM-2014-JCTC-Salvalaglio-Parrinello} 
to estimate the mean transition times between the metastable ($\alpha$) and the equilibrium ($\beta$) states 
of the peptide. WT-metaD was performed using both torsional angles $\Phi$ and $\Psi$  as CV.
We denote by $\tau$, the mean transition time over the barrier from the states, 
and by $\tau_M$, the mean transition time for the metadynamics run. The latter changes as the simulation 
progresses and is linked to the former through the acceleration factor 
$\alpha(t) \equiv \langle e^{\beta V(s,t)} \rangle_M = \tau/\tau_M(t)$, where the angular brackets 
$\langle \dots \rangle_M$ denote an average over a metadynamics run confined to the metastable basin, 
and $V(s,t)$ is the metadynamics time-dependent bias. To avoid depositing bias in the transition state region, 
we increase the time lag between two successive Gaussian depositions in the WT-metaD 
framework~\cite{SM-2013-PRL-Tiwary-Parrinello,SM-2014-JCTC-Salvalaglio-Parrinello} to $20$ ps and decrease the bias 
factor to $5$. The statistics for $\tau_{\alpha \to \beta}^{(\textrm{num})}$ and $\tau_{\beta \to \alpha}^{(\textrm{num})}$ 
conformed to a Poisson distribution with means $\mu_{\alpha \to \beta}= 5 \pm 2$~ns and $\mu_{\beta \to \alpha} = 125 \pm 37$~ns
and variance $\lambda_{\alpha \to \beta}=6$~ns and $\lambda_{\beta \to \alpha} = 116$~ns, respectively . 
The statistics obey a two-sample Kolmogorov-Smirnov test~\cite{SM-2014-JCTC-Salvalaglio-Parrinello} with $p$-value 
equal to $0.81$ and $0.76$, respectively. 
This yields the numerical ratio $k^{(num)}_{\beta \to \alpha}/k^{(num)}_{\alpha \to \beta} = (4.0 \pm 1.5) \times 10^{-2}$.

\section{Linear DNA denaturation bubble}
The cooperative opening and closure of a sequence of DNA consecutive base-pairs (bps) is central in biological mechanisms. 
The associated characteristic times measured experimentally by Altan-Bonnet \textit{et al.}~\cite{SM-2003-PRL-Altan-Krichevsky} 
showed large bubble lifetimes of $20  - 100~\mu$s and nucleation time of several $m$s. 
We use the DNA model of Refs.~\onlinecite{SM-2013-PRE-Dasanna-Manghi,SM-2015-JCP-Sicard-Manghi}, 
where the mesoscopic DNA model consists in two interacting bead-spring chains each made of $N = 50$ beads 
(of diameter $a = 0.34$ nm) at position $\textbf{r}_i$, with a AT-rich region of $30$ bps in the middle, 
and a GC region of $10$ bps at each extremity. The Hamiltonian is $\mathcal{H} =\mathcal{H}_{el}^{(1)} 
+ \mathcal{H}_{el}^{(2)} + \mathcal{H}_{tor} + \mathcal{H}_{int}$, where the first two contributions are elastic 
energies of the strands $j=1,2$ which include both stretching and bending energies
\begin{equation}
\mathcal{H}_{el}^{(j)} = \sum_{i=0}^{N-1} \frac{\kappa_s}{2}(r_{i,i+1}-a_\textrm{ref})^2 
+ \sum_{i=0}^{N-1}\frac{\kappa_\theta}{2}(\theta_i-\theta_\textrm{ref})^2.
\end{equation}
The stretching modulus, $a^2\beta_0 \kappa_s = 100$, is a compromise between numerical efficiency 
and experimental values~\cite{SM-Hugel-PRL2005}, where $\beta_0^{-1} = k_B T_0$ is the thermal energy,
$T_0 = 300$ K is the room temperature, and $a_\textrm{ref}=0.357$ nm. The bending modulus is large, $\beta_0 \kappa_\theta = 600$, 
to maintain the angle between two consecutive tangent vectors along each strand $\theta_i$ to the 
fixed value $\theta_\textrm{ref} = 0.41$ rad. Each strand is thus modeled as a freely rotating chain (FRC)~\cite{SM-Grosberg-AIP1994}.
The third and fourth terms of $\mathcal{H}$ are the torsional energy and hydrogen-bonding interactions, respectively. 
The torsional energy is modeled by a harmonic potential
\begin{equation}
\mathcal{H}_{tor} = \sum_{i=0}^{N-1} \frac{\kappa_{\phi,i}}{2}(\phi_i-\phi_\textrm{ref})^2 ,
\end{equation}
where $\phi_i$ is defined as the angle between two consecutive base-pair vectors
$\brho_i \equiv \textbf{r}_i^{(1)}-\textbf{r}_i^{(2)}$ and  $\brho_{i+1}$ ($\phi_\textrm{ref} = 0.62$ rad). 
The stacking interaction between base pairs is modeled through a $\kappa_{\phi,i}$ that depends on 
the value of the \textit{bare} dsDNA torsional modulus $\kappa_\phi$, and the distances between complementary bases,
$\kappa_{\phi,i} = \kappa_\phi [1-f(\rho_i)f(\rho_{i+1})]$, where 
\begin{equation} 
f(\rho_i) = \frac{1}{2}\Big[1+\erf\Big(\frac{\rho_i -\rho_b}{\lambda'}\Big)\Big], 
\label{stacking}
\end{equation}
and $\rho_i =|\brho_i|$. Hence, $\kappa_{\phi,i} = \kappa_\phi$ in the dsDNA state and $\kappa_{\phi,i} = 0$ 
in the ssDNA one. The actual values in the dsDNA state after equilibration, $\kappa^*_{\phi,\rm ds}$, 
are however different from the prescribed values, $\kappa_{\phi}$, due to thermal fluctuations and non-linear potentials 
entering the Hamiltonian. 
The hydrogen-bonding interaction is modeled by a Morse potential
\begin{equation}
\mathcal{H}_{int} = \sum_{i=0}^{N-1} A (e^{-2\frac{\rho_i-\rho_\textrm{ref}}{\lambda}} -2e^{-\frac{\rho_i-\rho_\textrm{ref}}{\lambda}}) ,
\end{equation}
where $\rho_\textrm{ref}=1$ nm, $\lambda=0.2$ nm, and $\beta_0 A=8$ and $12$ for AT and GC bonding, respectively, 
as in Refs.~\onlinecite{SM-Dasanna-EPL2012, SM-2013-PRE-Dasanna-Manghi,SM-2015-JCP-Sicard-Manghi}.
The fitted values for the dsDNA persistence length and the pitch are $\ell_{\rm ds}\simeq160$~bps 
and $p = 12$~bps for the relevant range of $\beta_0\kappa_\phi$ we are interested in, which are comparable to 
the actual dsDNA values ($\ell_{\rm ds}\simeq150$~bps and $p= 10.4$~bps). The ssDNA persistence length 
is $\ell_{\rm ss} = 3.7$~nm, compatible with experimental measurement~\cite{SM-Tinland-Macro1997}, 
even though in the upper range of measured values.
The evolution of $\textbf{r}_i(t)$ is governed by the overdamped Langevin equation, 
integrated using a Euler's scheme,
\begin{equation}
\zeta \frac{d\textbf{r}_i}{dt} = -\nabla_{\textbf{r}_i}\mathcal{H}({\textbf{r}_j}) + \mathbf{\xi}(t) ,
\end{equation}
where $\zeta=3\pi\eta a$ is the friction coefficient for each bead of diameter $a$ with 
$\eta=10^{-3}$ Pa.s the water viscosity. 
The diffusion coefficient, $D_\textrm{diff} \equiv k_BT_0/3\pi\eta a$, thus takes into account 
the level of coarse-graining of the mesoscopic model involved in the kinetics associated 
to the smoothed free-energy landscape~\cite{SM-Murtola-PCCP2009}. 
The random force of zero mean $\mathbf{\xi}_i(t)$ obeys 
the fluctuation-dissipation relation $\langle \mathbf{\xi}_i(t).\mathbf{\xi}_i(t')\rangle =6k_BT\zeta\delta_{ij}\delta(t-t')$. 
Lengths and energies are made dimensionless in the units of $a=0.34$ nm and $k_BT_0$, respectively. 
The dimensionless time step is $\delta\tau = \delta t k_B T_0/(a^2\zeta)$, set to $5 \times 10^{-4}$ 
($\delta t=0.045$ ps) for sufficient accuracy~\cite{SM-Dasanna-EPL2012,SM-2013-PRE-Dasanna-Manghi,SM-2015-JCP-Sicard-Manghi}. 
This set of parameters induces zipping velocities $v \approx 0.2-2$ bp/ns, compatible with experimental 
measurements~\cite{SM-Bustamante-COSB2000}.\\
\begin{figure}[b]
\includegraphics[width=1.0 \textwidth, angle=-0]{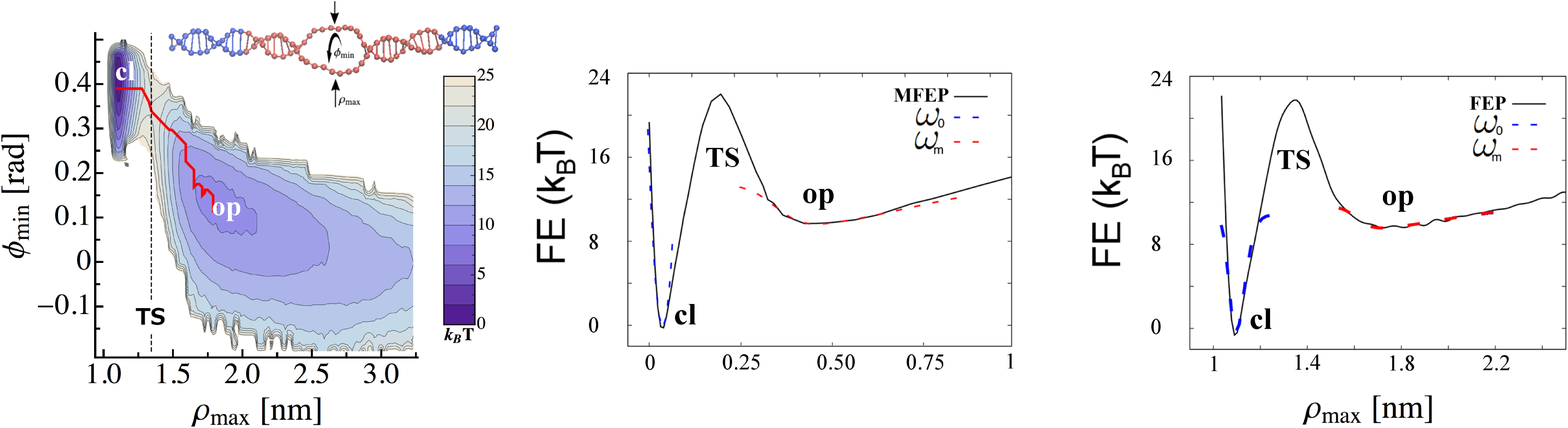}
 \caption{\textbf{Left panel} FE surface associated with the DNA bubble closure/nucleation mechanism projected 
 along the maximal distance between paired bases $\rho_{\textrm{max}}$ and the minimal twist angle between successive bps, 
 $\phi_{\textrm{min}}$ (see inset).  The contour lines are every two $k_B T$. The two stables basins associated with the opened (op) and closed (cl) states of the DNA bubble and the typical MFEP obtained within the steepest descent framework~\cite{SM-2013-JCP-Chen-Xiao} 
 are shown (red). 
\textbf{Middle panel} FE of the DNA bubble as a function of the progression along the typical MFEP (normalized to unity) 
obtained within the steepest descent framework~\cite{SM-2013-JCP-Chen-Xiao}. The nonlinear least-squares Marquardt-Levenberg algorithm was implemented to fit the parameters $\omega_0$ and $\omega_m$, measured in the equilibrium and metastable states, respectively.
\textbf{Right panel} FE profile of the system along $\rho_{\textrm{max}}$ reconstructed within the WT-metaD framework. 
The nonlinear least-squares Marquardt-Levenberg algorithm was implemented to fit the parameters  
$\omega_0$ and $\omega_m$, measured in the equilibrium and metastable states, respectively.
}
\label{figSM2}
\end{figure}

To reconstruct the FE surface, we performed WT-metaD coarse-grained simulations with the width of the DNA bubble, 
$\rho_{\max}(t)$, as CV using the version 2.3 of the plugin for free-energy calculation, named PLUMED \cite{SM-2014-CPC-Tribello-Bussi}
According to the algorithm introduced by Barducci \textit{et al.}~\cite{SM-2008-PRL-Barducci-Parrinello,SM-2009-JCC-Bonomi-Parrinello} 
a Gaussian is deposited every $25$ ps with intial height of $0.1\,k_B T$ and a bias factor of
$5$ at $T=300$ K. The resolution of the recovered free-energy landscape is determined by the width 
of the Gaussians $\sigma = 0.1$ in units of the CV. 
As described in previous work~\cite{SM-2015-JCP-Sicard-Manghi}, we put a wall 
at $\rho_{\max} \approx 4$ nm to prevent the system to escape from the metastable state
(and therefore entering in the zipping regime, \textit{i.e.} a far from equilibrium 
process~\cite{SM-Dasanna-EPL2012, SM-2013-PRE-Dasanna-Manghi}). 
We checked that a slight change in the position of the wall ($\rho_{\max}=3.5,4,4.5,5.5,7$ nm) does not change significantly 
the results, particularly the positions of the local minimum and the saddle, as well as the barrier height. 
To explore the \textit{slow} entropic contribution associated to the DNA bubble metastable basin we chose to follow the 
evolution of the minimal twist angle $\Phi_{\textrm{min}}$ inside the bubble~\cite{SM-2015-JCP-Sicard-Manghi}, as shown in the inset 
in Fig.~\ref{figSM2} (left panel), reconstructed afterwards using the \textit{reweighting technique} 
of Bonomi et al.~\cite{SM-2009-JCC-Bonomi-Parrinello}.\\

The analysis of the FE surface associated with the bubble closure and opening mechanisms, 
as shown in Fig.~\ref{figSM2} (left panel), allowed us to determine the value the FE of formation,  
$\Delta F^0 = F(op) - F(cl) = 9.0 \pm 0.1~k_B T$ and the activation energies, 
$\Delta F_{cl \to op} = 21.8 \pm 0.1~k_B T$ 
and $\Delta F_{op \to cl} = 12.9 \pm 0.1~k_B T$ along the MFEP obtained within the steepest descent framework~\cite{SM-2013-JCP-Chen-Xiao}, 
as shown in Fig.~\ref{figSM2} (middle panel), The FE of formation, $\Delta F^* = 6.7 \pm 0.1~k_B T$, 
defined in term of the probability distribution of $\rho_{\textrm{max}}$ and $\Phi_{\textrm{min}}$, 
was computed considering the successive isosurfaces in the FE basins depicted in Fig.~\ref{figSM2} (left panel) 
as integration domains (cf. Eq.~(13) in the main text). 
In Fig.~\ref{figSM2} (middle panel), we show the FE of the system as a function of the progression along 
the typical MFEP (normalized to unity). The nonlinear least-squares Marquardt-Levenberg algorithm was implemented 
to fit the parameters $\omega_0$ and $\omega_{m}$ with Gaussian or skew-Gaussian distributions depending 
on the symmetric or asymmetric nature of the FE profile, respectively. We obtained  
$\omega_{m} = 5.3 \pm 0.2$ and $\omega_0 = 64.2 \pm 2.1$ for the metastable ($cl$) and equilibrium ($op$) basins, respectively.
Considering the Rouse model~\cite{SM-2005-JPCM-Ambjornsson-Metzler} valid for flexible polymer chain, 
the effective friction coefficient, $\gamma$, in Eq.~$16$ in the main text depends on the number of opened bps, 
$N_{\textrm{bub}}$, in the DNA bubble. The typical size observed in the simulations, $N_{bub}\approx 10$ bps, 
yields the relation $\gamma_{op}/\gamma_{cl} \approx N_{\textrm{bub}}$ between the effective frictions.
We obtain the transition rate ratio, $k_{cl \to op}/k_{op \to cl} = (1.5 \pm 0.6) \times 10^{-3}$.

In Fig.~\ref{figSM2} (right panel), we show the FE profile of the system along $\rho_{\textrm{max}}$ reconstructed within 
the WT-metaD framework. We determined the value of the FE of formation, $\Delta F^0 = 10.3 \pm 0.1~k_B T$ 
and the activation energies, $\Delta F_{cl \to op} = 22.6 \pm 0.1~k_B T$ 
and $\Delta F_{op \to cl} = 12.3 \pm 0.1~k_B T$. The nonlinear least-squares Marquardt-Levenberg algorithm was implemented 
to fit the parameters $\omega_0$ and $\omega_{m}$ with Gaussian distribution. We obtained  
$\omega_{m} = 5.4 \pm 0.4$ and $\omega_0 = 64.3 \pm 1.9$ for the metastable ($\alpha$) and equilibrium ($\beta$) basins, respectively.
The \textit{standard} KT yields $k^{(st)}_{cl \to op}/k^{(st)}_{op \to cl} = (4.0 \pm 0.7) \times 10^{-3}$.

We extended the Metadynamics scope~\cite{SM-2010-JCP-Xin-Hamelberg,SM-2013-PRL-Tiwary-Parrinello,SM-2014-JCTC-Salvalaglio-Parrinello} 
to estimate the mean transition times between the metastable ($op$) and the equilibrium ($cl$) states 
of the DNA bubble. WT-metaD was performed using the width $\rho_{\textrm{max}}$ as CV. 
Unlike in the FE surface reconstruction, no wall was added along the CV $\rho_{\textrm{max}}$ in that case.
We denote by $\tau$, the mean transition time over the barrier from the states, 
and by $\tau_M$, the mean transition time for the metadynamics run. To avoid depositing bias in the transition state region, 
we increase the time lag between two successive Gaussian depositions in the WT-metaD 
framework~\cite{SM-2013-PRL-Tiwary-Parrinello,SM-2014-JCTC-Salvalaglio-Parrinello} to $700$ ps and decrease the bias 
factor to $3$. The statistics for $\tau_{op \to cl}^{(\textrm{num})}$ and $\tau_{cl \to op}^{(\textrm{num})}$ 
conformed to a Poisson distribution with means $\mu_{op \to cl}= 121 \pm 12~\mu$s and $\mu_{cl \to op} = 67 \pm 8$~ms
and variance $\lambda_{op \to cl}=110~\mu$s and  $\lambda_{cl \to op}=67$~ms, respectively . 
The statistics obey a two-sample Kolmogorov-Smirnov test~\cite{SM-2014-JCTC-Salvalaglio-Parrinello} with $p$-value 
equal to $0.86$ and $0.65$, respectively. 
This yields the numerical ratio $k^{(num)}_{cl \to op}/k^{(num)}_{op \to cl} = (1.8 \pm 0.4) \times 10^{-3}$.

\section{Circular DNA denaturation bubble}
\begin{figure}[h]
\includegraphics[width=1.0 \textwidth, angle=-0]{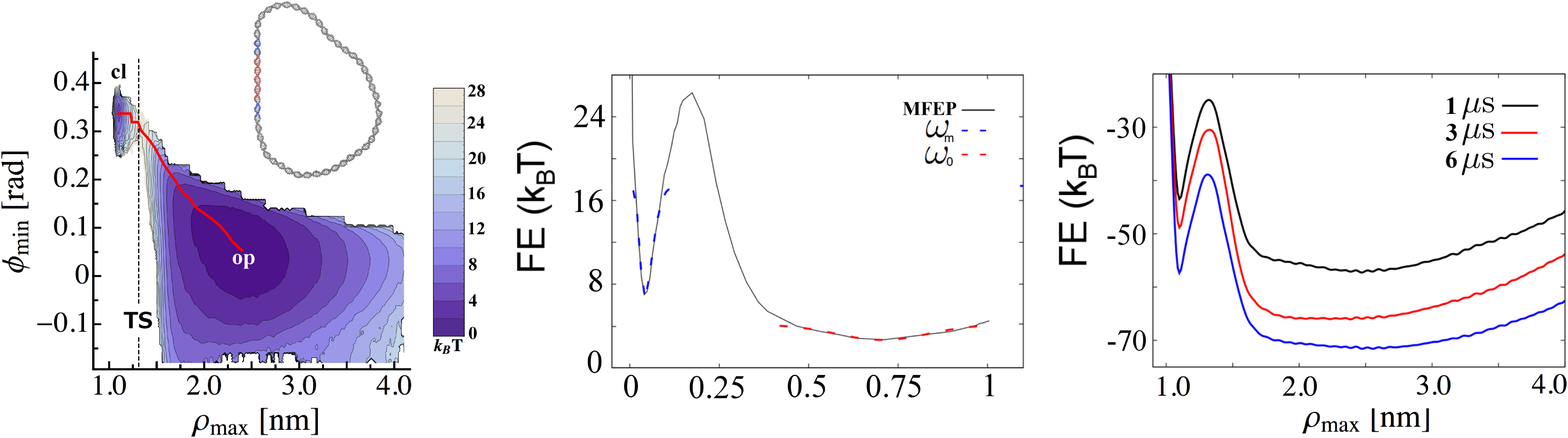}
 \caption{\textbf{Left panel} FE surface associated with the circular DNA bubble closure/nucleation mechanism projected 
 along the maximal distance between paired bases $\rho_{\textrm{max}}$ and the minimal twist angle between successive bps, 
 $\phi_{\textrm{min}}$.  The contour lines are every two $k_B T$. The two stables basins associated with the opened (op) and closed (cl) states of the DNA bubble and the typical MFEP obtained within the steepest descent framework~\cite{SM-2013-JCP-Chen-Xiao} 
 are shown (red). 
\textbf{Middle panel} FE of the circular DNA bubble as a function of the progression along the typical MFEP (normalized to unity) 
obtained within the steepest descent framework~\cite{SM-2013-JCP-Chen-Xiao}. The nonlinear least-squares Marquardt-Levenberg algorithm was implemented to fit the parameters $\omega_0$ and $\omega_m$, measured in the equilibrium and metastable states, respectively.
\textbf{Right panel} Temporal evolution of the FE profile of the system along $\rho_{\textrm{max}}$ reconstructed 
in the WT-metaD simulation. The convergence of the FE profile could not be achieved due to large entropic fluctuations.}
\label{figSM3}
\end{figure}

The circular DNA (cDNA) is described with the same DNA model used for the linear DNA, where the two single strands 
are modeled as freely rotating chains of $N=246$ beads of diameter $a=0.34$ nm with a AT-rich region of $30$ bps 
clamped by a closed circular GC region of $(N-30)$ bps. The size of these AT-rich regions was chosen 
so that it is larger than the size of the representative \textit{long-lived} denaturation bubbles studied in this work.
The dsDNA minicircle is described by a circular helix where a helical line of radius $\alpha$ coils around 
a torus of radius $R$ in the $x-y$ plane. The centers of the beads on each strand initially coincide with 
the surface of this torus in Cartesian space according to the equations
\begin{equation}
\left\{
\begin{aligned}
x_n^{(j)} &= \Big( \alpha \sin\Big(n\frac{2\pi}{p} + \psi^{(j)}\Big) + R \Big) \times \cos(n\theta) \\
y_n^{(j)} &= \Big( \alpha \sin\Big(n\frac{2\pi}{p} + \psi^{(j)}\Big) + R \Big) \times \sin(n\theta) \\
z_n^{(j)} &= \alpha \cos\Big(n\frac{2\pi}{p} + \psi^{(j)}\Big)
\end{aligned}
\right.
\end{equation}
with $x_n^{(j)}$, $y_n^{(j)}$ and $z_n^{(j)}$ the Cartesian coordinates of  bead $n$ on strand $j$. 
The parameter $\psi^{(1)}=0$ for the first strand and $\psi^{(1)}=\pi$ for the second strand. The cross-sectional 
radius $\alpha$ is set equal to half the equilibrium base-pair distance, $\rho_{\textrm{ref}} = 1$~nm, considered 
in previous work~\cite{SM-2013-PRE-Dasanna-Manghi,SM-2015-JCP-Sicard-Manghi}. 
The twist angle between two base-pairs is defined as $\phi =2\pi/p $, where $p=12.3$ is the DNA pitch, 
\textit{i.e.} the number of bps corresponding to one complete helix turn. For purposes of generating the initial 
conformations, the bending angle per axis segment between the centers of two consecutive bps 
is set initially at $\theta = 2\pi/N$.
We constrained a sequence of 10 GC bps on each extremity of the AT-rich region to be aligned arbitrarily 
along the Z-axis, as depicted in Fig.~\ref{figSM3} (left panel).
The superhelical densities $\sigma = \frac{Lk - Lk^0}{Lk^0} = \frac{\Delta Lk}{Lk^0}$ along with the sizes $N$ of the 
minicircles was specifically chosen to tune the value of the excess of linking number $\Delta L_k < 1$. 
The parameter $Lk = 20$ represents the linking numbers of the cDNA molecule  
and $Lk^0$ is defined as $Lk^0 = N/p_0$, with $p_0 = 12.0$ the equilibrium pitch  
measured in the \textit{open linear} states.\\

The analysis of the FE surface associated with the bubble closure and opening mechanisms, 
as shown in Fig.~\ref{figSM3} (left panel), allowed us to determine the value the FE of formation,  
$\Delta F^0 = F(op) - F(cl) = -4.4 \pm 0.5~k_B T$ and the activation energies, 
$\Delta F_{cl \to op} = 17.8 \pm 0.5~k_B T$ 
and $\Delta F_{op \to cl} = 23.5 \pm 0.4~k_B T$ along the MFEP obtained within the steepest descent framework~\cite{SM-2013-JCP-Chen-Xiao}, 
as shown in Fig.~\ref{figSM3} (middle panel). The FE of formation, $\Delta F^* = -8.6 \pm 0.4~k_B T$, 
defined in term of the probability distribution of $\rho_{\textrm{max}}$ and $\Phi_{\textrm{min}}$, 
was computed considering the successive isosurfaces in the FE basins depicted in Fig.~\ref{figSM3} (left panel) 
as integration domains. 
In Fig.~\ref{figSM3} (middle panel), we show the FE of the system as a function of the progression along 
the typical MFEP (normalized to unity). The nonlinear least-squares Marquardt-Levenberg algorithm was implemented 
to fit the parameters $\omega_0$ and $\omega_{m}$ with skew-Gaussian distributions due to the asymmetric nature of the FE shape. 
We obtained  $\omega_{m} = 69.5 \pm 3.1$ and $\omega_0 = 3.7 \pm 0.2$ for the metastable ($op$) and equilibrium ($cl$) basins, respectively.
Considering the Rouse model~\cite{SM-2005-JPCM-Ambjornsson-Metzler} valid for flexible polymer chain, 
the effective friction coefficient, $\gamma$, in Eq.~$16$ in the main text depends on the number of opened bps, 
$N_{\textrm{bub}}$, in the DNA bubble. The typical size observed in the simulations, $N_{bub}\approx 12$ bps, 
yields the relation $\gamma_{op}/\gamma_{cl} \approx N_{\textrm{bub}}$ between the effective frictions. 
We obtain the transition rate ratio, $k_{cl \to op}/k_{op \to cl} = (1.0 \pm 0.4) \times 10^{6}$.

In Fig.~\ref{figSM3} (right panel), we show the temporal evolution of the FE profile of the system 
along $\rho_{\textrm{max}}$ reconstructed in the WT-metaD simulation. In such case, the
convergence of the FE profile could not be achieved due to large entropic fluctuations.
However, the analysis of the converged FE surface obtained in Fig.~\ref{figSM3} (left panel) was achievable 
with the appropriate use of the auxiliary variable $\Phi_{\textrm{min}}$.

We extended the Metadynamics scope~\cite{SM-2010-JCP-Xin-Hamelberg,SM-2013-PRL-Tiwary-Parrinello,SM-2014-JCTC-Salvalaglio-Parrinello} 
to estimate the mean transition times between the metastable ($op$) and the equilibrium ($cl$) states 
of the DNA bubble. WT-metaD was performed using the width $\rho_{\textrm{max}}$  as CV.
The statistics for $\tau_{cl \to op}^{(\textrm{num})}$ conformed to a Poisson distribution with means 
$\mu_{op \to cl}= 4.9 \pm 0.6$ ms  and variance $\lambda_{op \to cl}=6.0$ ms. The statistics obeys a two-sample 
Kolmogorov-Smirnov test~\cite{SM-2014-JCTC-Salvalaglio-Parrinello} with $p$-value equal to $0.71$. 
However, the numerical estimation of the transition time $\tau_{cl \to op}^{(\textrm{num})}$ was not achievable
within the metaD framework, as the shape of the original FE surface could not be evenly maintained after
the addition of the bias potential due to \textit{large} entropic fluctuations.
Nevertheless, our approach allowed us to asses the transition rate ration and to estimate $\tau_{cl \to op} = 80 \pm 40$ min.

\end{document}